\documentclass[conference]{IEEEtran}
\IEEEoverridecommandlockouts
\usepackage{cite}
\usepackage{amsmath,amssymb,amsfonts}
\usepackage{algorithm}
\usepackage{algpseudocode}
\usepackage{graphicx}
\usepackage{textcomp}
\usepackage{xcolor}
\usepackage{tabularx}
\usepackage{booktabs}
\def\BibTeX{{\rm B\kern-.05em{\sc i\kern-.025em b}\kern-.08em
    T\kern-.1667em\lower.7ex\hbox{E}\kern-.125emX}}

\usepackage{pgfplots}
\pgfplotsset{compat=1.18}

\usepackage[hidelinks]{hyperref}

\usepackage[switch]{lineno}

\usepackage{tikz}
\usetikzlibrary{shapes.geometric, arrows.meta, positioning, fit, backgrounds, calc}
\tikzset{
  agent/.style={
    draw, rounded corners=3pt, align=center,
    minimum height=1.05cm, minimum width=2.6cm,
    fill=gray!9, font=\small\sffamily, line width=0.7pt
  },
  io/.style={
    draw, trapezium,
    trapezium left angle=72, trapezium right angle=108,
    align=center, minimum height=0.85cm, minimum width=1.5cm,
    font=\small\sffamily, line width=0.7pt
  },
  terminal/.style={
    draw, rounded corners=10pt, align=center,
    minimum height=0.85cm, minimum width=2.0cm,
    fill=green!10, font=\small\sffamily, line width=0.7pt
  },
  decision/.style={
    draw, diamond, aspect=1.9, align=center,
    inner sep=1pt, font=\small\sffamily, line width=0.7pt
  },
  arr/.style   = {->, >=Stealth, line width=0.85pt},
  darr/.style  = {->, >=Stealth, dashed, line width=0.75pt},
  darrSpec/.style = {darr, teal!70!black},
  darrFH/.style   = {darr, blue!65!black},
  darrRTL/.style  = {darr, orange!75!black},
  darrCEX/.style  = {darr, red!60!black},
  lbl/.style  = {font=\scriptsize\sffamily},
}

\begin{document}

\title{Closing the Loop on LLM-Generated RTL Assertions with Quality-Aware Formal Verification\\

}

  \author{
  \IEEEauthorblockN{Ramesh Krishnamurthy, Danial Chitnis, Themis Prodromakis}
  \IEEEauthorblockA{
  School of Engineering, The University of Edinburgh, Edinburgh, United Kingdom\\
  \href{mailto:email@domain.com}{rkrishn3@ed.ac.uk}
  }
  }


\maketitle

\begin{abstract}
Large language model (LLM) based assertion generation is making formal verification more accessible for Register Transfer Level (RTL) designs, but three practical issues remain. Generated properties can pass for the wrong reason, proof cost can vary widely from one design to another, and failing traces are hard to interpret. This paper presents a lightweight, open-source framework that addresses these issues in one loop. Our method combines mutation-guided refinement to reject weak assertions, including vacuous ones and those that fail to distinguish faulty behavior, a solver-selection stage that chooses among candidate Satisfiability Modulo Theories (SMT) backends using RTL structure, and causal narrative synthesis to explain why a proof failed. Across diverse RTL designs, the framework improves confidence in generated assertions, reduces runtime variability over fixed-solver choices, and produces failure explanations that remain grounded in the counterexample trace. The results suggest that quality-aware closure, rather than assertion generation alone, is the missing step for practical LLM-assisted formal verification.
\end{abstract}


\section{Introduction}
Formal verification (FV) offers a rigorous alternative to simulation for proving hardware correctness, but its deployment remains constrained by the effort needed to write high-quality assertions and configure proof backends effectively. Recent LLM-based frameworks have reduced the front-end burden of assertion generation, including systems that translate specifications into assertions and systems that use both the design specification and the RTL code as input \cite{specllm,assertllm,assertionforge}. More broadly, recent surveys and agentic workflows show growing interest in LLM-assisted hardware design and verification, but they also highlight the need for stronger quality control and evaluation tied to actual proof results from formal tools \cite{hwdv_survey,hey_ai}. Generation alone, however, is not enough. A generated assertion may pass vacuously, run slowly when paired with an unsuitable solver backend, or fail with a trace that is hard to understand.

We target these three bottlenecks directly. First, vacuity remains a known threat in formal verification because properties can hold for trivial reasons rather than because the intended behavior was actually exercised \cite{beer2001vacuity}. Second, backend selection remains highly instance-dependent; prior work on SMT formulas and word-level netlists shows that structural features can strongly affect solver performance \cite{machsmt,btor2select}. Third, raw counterexample traces remain difficult to interpret, especially for users who are new to FV, because long traces often contain many signal changes and the first divergence does not necessarily coincide with the cycle at which the assertion actually fails.

Recent related work covers important parts of this picture, but not the whole loop. Assertion-generation systems such as AssertLLM, Spec2Assertion, AssertionForge, and DeepAssert improve how properties are synthesized from text, structure, or module context \cite{assertllm,spec2assert25,assertionforge,deepassert25}. Recent work has also coupled LLM-generated properties with mutation testing to assess invariant quality \cite{date24_mutation}. In our flow, mutation serves a different role: mutant outcomes trigger \textit{verification harness refinement} and help decide whether the harness is strong enough to proceed. This distinction is deliberate. Assessing invariant quality is useful as a diagnostic, but our goal is closure: when mutants survive, the framework uses that evidence to strengthen the harness or to recognize that a survivor remains unobservable within the explored proof window. The two strategies are therefore complementary: quality assessment can score a generated property set, whereas mutation-guided refinement uses the same mutation evidence as a control signal for refinement. Agentic workflow systems such as Saarthi show how LLM agents can orchestrate formal tasks \cite{saarthi24}, but the quality-control layer, i.e., checking assertion meaningfulness, selecting a backend, and explaining failures from trace evidence, remains the gap addressed here.

This paper presents a quality-aware framework for LLM-generated RTL assertions. The framework combines: (i) \emph{Mutation-Guided Refinement (MGR)} to reject weak assertions, including vacuous ones and those that fail to distinguish faulty behavior, 
(ii) an \emph{architecture-aware solver selector} that uses RTL structure before the design is translated into solver-specific forms, and (iii) \emph{Causal Narrative Synthesis (CNS)} to summarize failing traces in human-readable form. The contribution is a compact loop that improves confidence in generated assertions, stabilizes runtime, and makes failures easier to debug in an \textit{open-source} flow built on Yosys-based tooling \cite{yosys}.\footnote{Code and benchmark artifacts are available at \href{https://github.com/eelab-dev/EEdigits}{\texttt{github.com/EEdigits}}.}

\section{Framework}

\begin{figure*}[t]
\centering
\resizebox{\textwidth}{!}{%
\begin{tikzpicture}

  \node[io]       (spec)    at ( 0.0, 0) {Spec.};
  \node[io]       (rtl)     at ( 2.5, 0) {RTL};
  \node[agent, dashed, fill=gray!12]
                  (a1)      at ( 6.0, 0) {\textbf{I}\\[1pt]Initial\\Harness};
  \node[io]       (smteng)  at (10.5, 0) {SMT\\Backend};
  \node[agent]    (a3)      at (15.5, 0) {\textbf{III}\\[1pt]Mutation-Guided\\Refinement};
  \node[decision] (d1)      at (19.5, 0) {Kill rate\\${\geq}\,T$\,?};
  \node[decision] (d2)      at (23.5, 0) {Proof\\result?};

  \node[agent]    (a2)   at (10.5, 3.2)
                  {\textbf{II}\\[1pt]SMT Solver\\Selection\\[2pt]{\scriptsize(run once)}};

  \node[terminal] (ver)  at (23.5, -2.2) {RTL\\Verified};
  \node[agent]    (a4)   at (27.5,  0.0) {\textbf{IV}\\[1pt]Causal Narrative\\Synthesis};
  \node[io]       (rep)  at (27.5, -2.2) {Failure\\Report};

  \draw[arr] (spec) -- (rtl);
  \draw[arr] (rtl)  -- (a1);
  \draw[arr] (a1)   -- (smteng)
      node[midway, above, lbl, align=center] {verification\\harness};
  \draw[arr] (smteng) -- (a3);
  \draw[arr] (a3)   -- (d1);
  \draw[arr] (d1)   -- (d2)
      node[midway, above, lbl] {Yes};
  \draw[arr] (d2.south) -- (ver)
      node[midway, right, lbl] {pass};
  \draw[arr] (d2.east)  -- (a4)
      node[midway, above, lbl] {fail};
  \draw[arr] (a4.south) -- (rep);

  \draw[arr]
      (spec.north) -- ++(0, 1.4) -| (a1.north)
      node[pos=0.45, above, lbl] {properties};

  \coordinate (rtlL) at ($(rtl.north)+(-0.15,0)$);
  \draw[arr]
      (rtlL) -- (rtlL |- a2.west) -- (a2.west)
      node[pos=0.75, above, lbl] {structural metrics};

  \draw[arr]
      (a2.south) -- (smteng.north)
      node[midway, right, lbl] {solver cfg};

  \draw[darrSpec]
      ($(rtl.north)+(+0.15,0)$) -- ++(0, 4.0) -| (a3.north)
      node[pos=0.4, above, lbl, text=teal!70!black] {RTL source};

  \draw[arr]
      (d1.south) -- ++(0, -2.0) -| (a1.south)
      node[pos=0.07, right=3pt, lbl, fill=white, inner sep=1pt] {No}
      node[pos=0.42, below, lbl] {survived mutants + audit evidence};

  \coordinate (NL) at ($(a4.north west)!0.18!(a4.north east)$);
  \coordinate (NC) at (a4.north);
  \coordinate (NR) at ($(a4.north west)!0.82!(a4.north east)$);

  \draw[darrFH] ($(NL)+(0,1.1)$) -- (NL)
      node[pos=0, above, font=\scriptsize\sffamily, align=center, text=blue!65!black]
          {Verification\\Harness};
  \draw[darrRTL] ($(NC)+(0,1.1)$) -- (NC)
      node[pos=0, above, font=\scriptsize\sffamily, align=center, text=orange!75!black]
          {RTL};
  \draw[darrCEX] ($(NR)+(0,1.1)$) -- (NR)
      node[pos=0, above, font=\scriptsize\sffamily, align=center, text=red!60!black]
          {CEX\\Trace};

\end{tikzpicture}
}
\caption{The proposed quality-aware verification loop. The initial harness stage is shown only as a bootstrap. Solver selection is computed once from RTL structure and reused during mutation-guided refinement. The paper's technical contributions begin with solver selection and continue through mutation-guided refinement and concise failure narratives.}
\label{fig:Agentic FV}
\end{figure*}
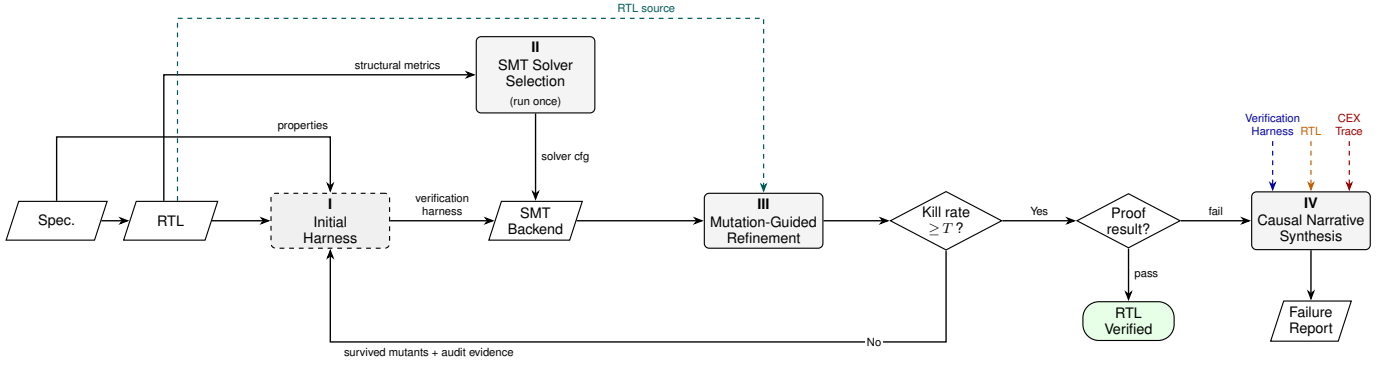

Figure~\ref{fig:Agentic FV} summarizes the proposed flow. Given an RTL design and a natural-language specification, an LLM first produces an initial verification harness. In this paper, a verification harness means the formal assertions, assumptions, and any helper logic used to verify the RTL design; we refer to it hereafter simply as the \textit{harness}.
The framework then profiles RTL structure once, selects a solver backend, and emits the proof configuration used in subsequent audit runs. MGR injects synthetic RTL faults and checks whether the current assertions detect them under that fixed configuration. Surviving mutants trigger harness refinement. Once the assertion set kills the relevant mutants at an acceptable rate, the final proof outcome is recorded; if the proof fails, the resulting counterexample is translated into a short root-cause narrative. The front-end harness stage is included for completeness, but the technical contributions begin after that point: MGR, solver selection, and CNS. In the present experiments, the initial harness is generated with RTL context because the study starts from existing public RTL designs and treats harness generation as a bootstrap step rather than a contribution under evaluation. Most of the quality-control loop is tool-driven: the mutation audit and kill-rate computation in MGR, together with RTL profiling and solver selection, are \textit{deterministic} stages, while LLM calls are used only for harness construction/refinement and for natural-language failure summaries.

\subsection{Mutation-Guided Refinement}
The goal of MGR is to \emph{test the test}~\cite{mutation_jia}. Starting from the original RTL $D$ and the current assertion set $\mathcal{A}$, the framework injects faults into $D$ and checks whether $\mathcal{A}$ still passes. Any passing mutant reveals a weakness in the harness; the survived mutant is then fed back to the LLM for the next refinement round. Vacuity is one failure mode of weak assertions, but MGR addresses a broader problem: whether the current harness is sufficiently discriminative under injected faulty behaviors. Satisfiability checks may be used to detect strict vacuity, while MGR strengthens properties that still pass but fail to distinguish relevant mutants.

Unlike prior mutation-based quality checks for generated properties \cite{date24_mutation,deepassert25}, MGR uses mutant outcomes to decide whether the harness should be accepted, refined again, or interpreted (cautiously) when a surviving mutant does not produce an observable failure within the explored proof window. For a mutant set $M$, we define the effective kill rate as:
\begin{equation}
\rho_{\mathrm{eff}} = \frac{\#\mathrm{Killed}}{\#\mathrm{Total} - \#\mathrm{Invalid}}.
\end{equation}
In the current workflow, we set the refinement trigger to $T = 85\%$: if the effective kill rate falls below this value, another refinement round is started. This threshold is a practical floor rather than a final quality target; in our experiments, robust harnesses typically converged to 90--100\% effective kill on non-invalid mutants. If the effective kill rate remains low, surviving mutants are fed back to the LLM for strengthening, turning formal verification into an \textit{adversarial refinement loop} rather than a one-shot generation task.

\paragraph*{MGR workflow}
Each mutant is verified independently under the same harness and proof configuration. Results are classified as \textsc{Killed} (assertions detect the mutant), \textsc{Survived} (missed by the harness), \textsc{Invalid} (the edit no longer produces a meaningful hardware variant), or \textsc{Timeout}/\textsc{Inconclusive}. These per-mutant outcomes and aggregate audit results form the audit evidence returned to the refinement stage. Surviving mutants are fed back to the LLM to strengthen the next harness round. \textsc{Invalid} mutants are excluded from the denominator of $\rho_{\mathrm{eff}}$ because they do not exercise the harness.

\subsection{Architecture-Aware Solver Selection}
To reduce runtime unpredictability, we profile each design before translation into solver-specific forms. We compute a lightweight structural score:
\begin{equation}
\chi = (D_A \times W) + D_M + \frac{D_R}{I_C},
\end{equation}
where $W$ is dominant datapath width, $D_A$ is arithmetic density, $D_M$ is mux density, $D_R$ is total memory size in bits, and $I_C$ is index complexity. The terms are normalized to a common reference scale and combined as a weighted sum for the final solver decision, preventing the memory term from dominating on small designs. Intuitively, high $D_M$ indicates control-steering logic, high $D_A \times W$ indicates wide datapaths, and high $D_R/I_C$ indicates memory-sensitive behavior. Unlike prior selectors that operate on SMT formulas or lowered Btor2 tasks \cite{machsmt,btor2select}, our selector works directly on RTL structure. This lets the flow choose a backend \textit{before} building the full proof instance, using signals that hardware engineers can understand and sanity-check.

\subsection{Causal Narrative Synthesis}
When verification fails, the framework extracts the failing property, a compact slice of the trace, and the most relevant signal transitions, then prompts the LLM to produce a concise explanation of the trigger condition and violated expectation. Counterexample explanation is not a new problem in formal verification; prior work has studied causality-based methods for highlighting why a trace violates a specification \cite{cex_causality}, and industrial studies show that explanation quality directly affects whether engineers can make effective use of formal verification results \cite{bosch_fv_explain}. Our focus here is different: CNS is evaluated as a \textit{faithfulness problem} rather than a fluency problem, so the explanation should point to the right cycle, the right signals, and the right failure mechanism without introducing claims that are not supported by the trace. This makes CNS a measurable \textit{debugging aid} rather than a presentation-only add-on. In the current workflow, the model is guided by the failing step and trace-based failure cues so that the explanation remains tied to the actual counterexample.

\section{Experimental Evaluation}
\label{sec:eval}
\subsection{Setup}

\begin{figure}[ht]
\centering
\begin{tikzpicture}
\begin{axis}[
    width=0.85\columnwidth,
    height=5.2cm,
    xlabel={Refinement Round},
    ylabel={Kill Rate (\%)},
    xmin=0.5, xmax=3.5,
    ymin=0, ymax=110,
    xtick={1,2,3},
    ytick={0,20,40,60,80,100},
    legend pos=south east,
    ymajorgrids=true,
    grid style=dashed,
    tick label style={font=\scriptsize},
    label style={font=\scriptsize},
    legend style={font=\tiny, cells={anchor=west}},
    nodes near coords,
    every node near coord/.append style={font=\tiny, yshift=2pt}
]

\addplot[color=black, mark=o]
    coordinates {(1,25)(2,35)(3,90)};
    \addlegendentry{UART\_TX}

\addplot[color=blue, mark=square*]
    coordinates {(1,10)(2,100)};
    \addlegendentry{I2C}

\addplot[color=red, mark=triangle*, every node near coord/.append style={yshift=-11pt}]
    coordinates {(1,70)(2,90)};
    \addlegendentry{VGA}

\end{axis}
\end{tikzpicture}
\caption{Illustrative MGR refinement trajectories for UART\_TX, I2C, and VGA; Table~\ref{tab:mgr} gives the cross-design summary.}

\label{fig:mgr}
\end{figure}
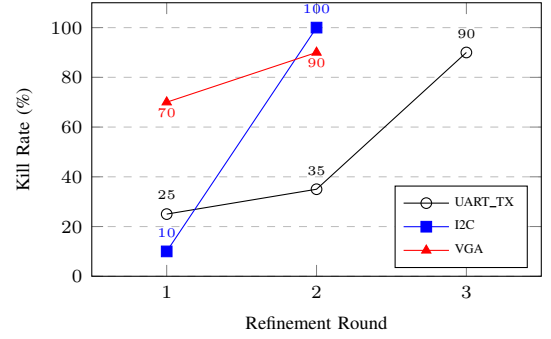

The framework is implemented on top of a white-box Yosys/SymbiYosys flow \cite{yosys}. An orchestration layer extracts structural metrics, selects a solver, and emits the SymbiYosys configuration automatically. In our current artifact, initial harness generation uses GPT~5.4 and CNS uses Claude Sonnet~4.6 (Anthropic). Each MGR campaign uses 20 mutants per round. We evaluate MGR on seven RTL designs: UART\_RX, UART\_Full, UP8, SHA3, SDRAM, VGA, and FFT. Fig.~\ref{fig:mgr} shows three supplementary learning curves: UART\_TX, I2C, and VGA. For solver selection, we use Yices, Bitwuzla, Z3, and CVC5 as representative SMT backends \cite{yices,bitwuzla,z3,cvc5}. We profile each design and report average step latency (ASL),
\begin{equation}
\mathrm{ASL} = \frac{T_{\mathrm{total}}}{k},
\end{equation}
where $T_{\mathrm{total}}$ is wall-clock proof runtime and $k$ is the bounded depth~\cite{biere99bmc}. Table~\ref{tab:rationale} lists the benchmark classes and chosen depths.

\begin{table}[t]
\caption{Benchmark classes and verification depths.}
\label{tab:rationale}
\centering
\small
\setlength{\tabcolsep}{3.5pt}
\begin{tabular}{@{}l l c@{}}
\toprule
\textbf{Class} & \textbf{Design} & \textbf{$k$} \\
\midrule
Protocol  & I2C Master  & 80  \\
Protocol  & UART (Full) & 120 \\
Memory    & Sync.\ FIFO & 15  \\
Memory    & SDRAM Ctrl  & 500 \\
Data      & FFT 256     & 80  \\
Data      & SHA3        & 80  \\
Peripheral & VGA LCD    & 80  \\
Control   & uP8 (ADD/ISA) & 45 \\
\bottomrule
\end{tabular}
\end{table}

\subsection{Mutation-Guided Refinement Results}
Table~\ref{tab:mgr} summarizes the best achieved MGR round across seven completed campaigns. The overall pattern is strong and consistent: six designs finish between 90\% and 100\% effective kill rate. UP8 and SHA3 reach 100\% in their first round. UART\_RX, UART\_Full, and VGA stop at 90\%, and FFT reaches 93.8\%. SDRAM is the only clear outlier. Its final score remains lower because some mutations change timing-related internal behavior, but within the bounded number of cycles (explored by the proof) they do not produce any observable failure, so the mutant survives even though the harness may not actually be weak. In other words, additional refinement stopped yielding new kills. We refer to that point as a \emph{plateau}.

\begin{table}[t]
\caption{Best MGR round per design (`R' = round).}
\label{tab:mgr}
\centering
\small
\begin{tabularx}{\columnwidth}{@{}l c c X@{}}
\toprule
\textbf{Design} & \textbf{Best} & \textbf{Eff. Kill} & \textbf{Observation} \\
\midrule
UART\_RX  & R1 & 90.0\%  & delay-only survivors \\
UART\_Full & R1 & 90.0\% & delay-only survivors \\
UP8       & R1 & 100.0\% & instruction-level checks \\
SHA3      & R1 & 100.0\% & fully specified datapath behavior \\
SDRAM     & R4 & 36.8\%  & no new kills after timing edits \\
VGA       & R2 & 90.0\%  & one-round refinement gain \\
FFT       & R2 & 93.8\%  & one residual inactive-branch edit \\
\bottomrule
\end{tabularx}
\end{table}

Across the suite, the most important cross-design pattern is the emergence of source-level edits that change the RTL text but not what the formal tool can observe. We refer to these as \emph{true equivalents}. Three categories recur repeatedly: delay annotations that disappear in RTL-level checking, inactive preprocessor branches, and timing-setting edits whose effect does not appear within the number of cycles explored by the proof. This distinction matters because a low plateau is not automatically evidence of a weak harness. In practice, MGR is most useful when it shows which survivors come from weak assertions and which come from mutations that the formal setup cannot observe.

\subsection{Solver Selection Results}
Table~\ref{tab:psp} reports solver performance across the current benchmark suite. The pattern is consistent with the proposed selector: control-heavy designs with small widths, such as I2C, VGA, and uP8, favor Yices, while the wide datapath benchmark SHA3
favors Bitwuzla. The table also shows why backend choice cannot be treated as a \textit{default setting}. Cases like FFT256, SHA3, and VGA, timeout under Z3 despite completing on other backends. The selector is therefore not chasing only small runtime
gains; it is avoiding bad choices that turn a solvable proof into a timeout. The memory-heavy uP8(ISA) case provides a useful calibration point: although its structure suggests an array-friendly backend (like Z3), Yices still wins at the tested proof bound. This shows that memory density alone is not enough to choose the solver, and that the selector must capture such \textit{hardware-specific tradeoffs}.

\begin{table}[t]
\caption{Normalized solver performance (ASL in s/step); T/O = timeout.}
\label{tab:psp}
\centering
\small
\setlength{\tabcolsep}{4pt}
\begin{tabular}{lcccc}
\toprule
\textbf{Benchmark} & \textbf{Bitwuzla} & \textbf{CVC5} & \textbf{Z3} & \textbf{Yices} \\
\midrule
I2C Master & 0.125 & 5.913 & 0.713 & \textbf{0.038} \\
UART (Full) & \textbf{0.008} & T/O & 0.017 & \textbf{0.008} \\
Sync.\ FIFO & \textbf{2.733} & T/O & 27.733 & 6.800 \\
SDRAM Ctrl & 0.026 & T/O & 0.290 & \textbf{0.024} \\
FFT 256 & 1.338 & 15.975 & T/O & \textbf{0.613} \\
SHA3 & \textbf{0.250} & 9.413 & T/O & 0.500 \\
VGA LCD & 0.388 & 8.725 & T/O & \textbf{0.150} \\
uP8 (ADD) & 3.556 & 44.000 & 0.156 & \textbf{0.044} \\
uP8 (ISA) & 0.667 & T/O & 18.756 & \textbf{0.556} \\
\bottomrule
\end{tabular}
\end{table}

To make the selector easier to interpret, Table~\ref{tab:baseline} compares our rule-based policy against fixed-solver baselines. ``Always Yices'' and ``Always Bitwuzla'' mean that the same solver is used  for every benchmark. ``Oracle'' is the unattainable upper bound that always picks the fastest solver after seeing the result.  On the current benchmark suite, the rule-based selector matches the oracle solver choice on every benchmark and reduces mean ASL by roughly 49\% compared with always using Yices.

\begin{table}[t]
\caption{Solver-selection baseline summary on the current suite.}
\label{tab:baseline}
\centering
\small
\begin{tabular}{lcc}
\toprule
\textbf{Policy} & \textbf{Mean ASL} & \textbf{Relative to Yices} \\
\midrule
Always Bitwuzla  & 1.010 & 1.04$\times$ \\
Always Yices     & 0.970 & 1.00$\times$ \\
Rule-based selector & \textbf{0.491} & 0.51$\times$ \\
Oracle best      & \textbf{0.491} & 0.51$\times$ \\
\bottomrule
\end{tabular}
\end{table}

\subsection{Causal Narrative Results}

We evaluated CNS on eight failed counterexamples drawn from four designs (FFT, VGA, UART\_RX, and SHA3). For each case, the ground truth consists of the manually established failing cycle, key signals, and failure mechanism. This ground truth was written before any CNS run and then verified in a second blind evaluation pass. Table~\ref{tab:cns} compares three settings: R1 baseline is the original zero-shot prompt, R2 refined is the shared improved prompt evaluated on six cases without case-specific hints, and R2 + hints adds such hints for the two hardest cases. The four binary scoring criteria answer the four basic debug questions: \emph{when} (correct failing cycle), \emph{where} (correct key signal), \emph{what kind} (correct mechanism class), and \emph{honest} (zero unsupported signal claims). A response is \textit{faithful} only when all four hold.

\begin{table}[t]
\caption{CNS faithfulness across prompt settings.}
\label{tab:cns}
\centering
\small
\begin{tabular}{lccc}
\toprule
\textbf{Metric} & \textbf{R1 baseline} & \textbf{R2 refined} & \textbf{R2 + hints} \\
\midrule
Correct cycle     & 4/8 & 6/6 & 8/8 \\
Correct signal    & 8/8 & 6/6 & 8/8 \\
Correct mechanism & 5/8 & 6/6 & 8/8 \\
Unsupported claims & 0  & 0   & 0   \\
Overall faithful  & 3/8 & $\geq$6/8 & 8/8 \\
\bottomrule
\end{tabular}
\end{table}

Even in the baseline, CNS identifies the correct signals in all eight cases and never invents signal names. Its errors are mainly about \emph{when} and \emph{how}: in longer traces, the baseline sometimes reports the first cycle where signals begin to diverge rather than the step at which the assertion fails, and it sometimes assigns a broad label such as ``timing error'' when the trace supports a more specific class. The refined prompt addresses both issues by stating the assertion-failure step explicitly and by giving clearer descriptions for each failure class. All six cases evaluated without case-specific hints are then faithful. The remaining two require extra hints, so the conservative claim is that refined CNS reaches at least 75\% faithfulness with the shared prompt, rising to 100\% with additional hints. Across all settings, signal identification is correct and no unsupported signal names are introduced. After refinement, CNS explanations remain short while preserving the key debug facts: responsible signal, failure type, and why the assertion fires.

\section{Discussion and Conclusion}
The results suggest that the central problem in LLM-assisted FV is not property generation alone, but \emph{quality-aware closure}. MGR is the strongest component: it directly addresses false confidence from weak properties, and six of seven designs reach 90--100\% effective kill rate. The SDRAM plateau illustrates an important structural limit: some timing-oriented mutations do not produce an observable failure within feasible proof depths, so they cannot be killed regardless of harness
quality. Recognizing that limit is itself a useful outcome of MGR. The solver selector provides a lightweight hardware-level rule for choosing the proof backend. CNS demonstrates that a zero-shot LLM reliably identifies the correct signals without hallucination, and a targeted two-part prompt refinement lifts overall faithfulness from 37\% to at least 75\%. Coverage remains an important metric in AI-for-FV research, but this paper focuses on a narrower and more defensible notion of quality: adversarial coverage through MGR, runtime stability through solver selection, and trace faithfulness through CNS.
Taken together, the results support a simple message: LLM-generated assertions become materially more \textit{trustworthy} when they are mutation-tested, structurally profiled, and paired with an open verification stack. Future work will add explicit coverage-closure metrics, compare the structural score ($\chi$) against additional solver-selection baselines, and extend the CNS evaluation to counterexamples from live regressions rather than mutation-triggered failures alone.

\section*{Acknowledgment}
The authors acknowledge UK Research and Innovation (UKRI) via its Engineering and Physical Sciences Research Council (EPSRC) funding for the AI Hub for Productive Research and Innovation in Electronics (APRIL), grant number EP/Y029763/1.Also, EDINA and ISG@University of Edinburgh for AI services.

\clearpage



\bibliographystyle{IEEEtran}
\bibliography{references}

\end{document}